# A Description of $^{200-204}Hg$ isotopes in the $SU(3) \leftrightarrow \overline{SU(3)}$ Transitional Region

M. A. Jafarizadeh$^{a,b}$ *, N. Fouladi$^c$, H. Sabri$^{c\dagger}$, P. Hosseinnezhade gavifekr $^c$, Z. Ranjbar $^c$

**a** Department of Theoretical Physics and Astrophysics, University of Tabriz, Tabriz 51664, Iran.

**b** Research Institute for Fundamental Sciences, Tabriz 51664, Iran.

**c** Department of Nuclear Physics, University of Tabriz, Tabriz 51664, Iran.

---

* E-mail: jafarizadeh@tabrizu.ac.ir
† E-mail: h-sabri@tabrizu.ac.ir

## Abstract

With employing the $SO(6)$ representation of eigenstates, the energy spectra and quadrupole transition rates of $SU(3) \leftrightarrow \overline{SU(3)}$ transition region are considered for systems with total boson number $N = 2, 3, 4$. The apparent level crossing and significant variations in the transition rates for $\chi = 0$, namely, the critical point of transitional Hamiltonian, suggest a phase transition between these limits. The parameter free (up to overall scale factors) predictions for spectra and $B(E2)$ rates by $\chi = 0$ are found to be in good agreement with those for nuclei provide empirical evidences for $O(6)$ dynamical symmetry, namely $^{200-204}Hg$ isotopes.

## Introduction

Nuclear transition regions have been considered the most complex and challenging of all nuclear regions. The investigation of significant changes in energy levels and electromagnetic transition rates resulting in the shape phase transitions [1-4] has received a lot of attention in recent years. The new symmetries called $X(5)$, $E(5)$ and $Z(5)$ are obtained within the framework of the collective model have employed to describe atomic nuclei at the critical points [5-8]. The parameter-free predictions provided by these symmetries are closely realized in some atomic nuclei.

In the interacting boson model (IBM) framework [9-12], a very simple two-parameter description has been used leading to a symmetry triangle describing many atomic nuclei. This model describes the nuclear structure of the even–even nuclei within the $U(6)$ symmetry, possessing the $U(5)$, $SU(3)$ and $O(6)$ dynamical symmetry limits. No phase transition is found between the $SU(3)$ and $O(6)$ vertices of the triangle. However, as discussed in Refs. [13-14] in the context of catastrophe theory, an analysis of the separatrix of the IBM-1 Hamiltonian in the coherent state formalism shows that there is a phase transition in between oblate and prolate deformed nuclei. This phase transition and its critical point symmetry, which in fact, coincides by the $O(6)$ limit have been described from the standpoint of physical observables in Refs.[15-17] by Jolie *et al*.

In this paper, we have considered the energy spectra and quadrupole transition rates for systems which localized in the $SU(3) \leftrightarrow \overline{SU(3)}$ transitional region in the IBM framework. With using the $SO(6)$ representation [18-21] for eigenstates, the matrix elements of quadrupole term in Hamiltonian have determined for systems with total boson number $N = 2,3,4$. The variation of control parameter, i.e. $\chi$, between limits, namely from $\chi = \sqrt{7}/2$ for $SU(3)$ to $\chi = -\sqrt{7}/2$ which correspond to $\overline{SU(3)}$ limit, suggest level crossing and also significant changes in transition rates. These results propose a phase transition in this region where $\chi = 0$, which is correspond to $O(6)$ dynamical symmetry limit, regards as critical point of this phase transition. Also, with employing a parameter free method in describing the considered quantities, the predictions of this approach for $\chi = 0$, have compared with the available experimental data for some nuclei which known as empirical evidences for $O(6)$ dynamical symmetry limit, namely $^{200-204}Hg$ isotopes while an acceptable degree of agreement is achieved.

This paper is organized as follows: section 2 briefly summarizes the theoretical aspects of considered Hamiltonian and $SO(6)$ representations for eigenstates, determined matrix elements of quadrupole term in Hamiltonian and also $B(E2)$ transition rates. Numerical results included some comparison between theoretical results with experimental data are presented in Section 3. Section 4 is devoted to summarize and some conclusion based on the results given in section 3.

## 2. The model
### 2.1. Transitional Hamiltonian and $SO(6)$ representation

The phase transitions have been studied widely in Refs.[15-17], are those of the ground state deformation. In the Interacting Boson Model (IBM), one would achieve a very simple two parameters description leading to a symmetry triangle which is known as extended Casten triangle [12]. There are four dynamical symmetries of the IBM called $U(5)$, $O(6)$, $SU(3)$ and $\overline{SU(3)}$ limits. They correspond to vibrational nuclei with a spherical form, namely $U(5)$, an axially symmetric prolate rotor with a minimum in the energy at $\gamma = 0°$ which corresponds to $SU(3)$ and an axially symmetric oblate rotor with a minimum at $\gamma = 60°$, namely $\overline{SU(3)}$. The fourth symmetry is located in the middle of the $SU(3) \leftrightarrow \overline{SU(3)}$ transitional region and corresponds to a rotor with a flat potential in $\gamma$, it means $O(6)$ limit, as have presented in Figure1. It is parameterized using the simple Hamiltonian [15-16]

$$\hat{H}(N,\eta,\chi) = \eta \hat{n}_d + \frac{\eta-1}{N}\hat{Q}_\chi \hat{Q}_\chi \quad , \tag{2.1}$$

Where $\hat{n}_d = d^\dagger \cdot \tilde{d}$ is the $d$-boson number operator and $\hat{Q}_\chi = (s^\dagger \tilde{d} + d^\dagger s)^{(2)} + \chi(d^\dagger \times \tilde{d})^{(2)}$ represents the quadrupole operator and $N(=n_s+n_d)$ stands for the total number of bosons. The $\eta$ and $\chi$ regard as control parameters while vary within the range $\eta \in [0,1]$ and $\chi \in [-\sqrt{7/2},+\sqrt{7/2}]$. Our considered region, namely, the prolate-oblate transitional region, passing through the $O(6)$ dynamical symmetry limit, is known to be situated close to the upper right leg of the extended Casten triangle with $\eta = 0$. In the following, we have employed the $SO(6)$ representation to determine the eigenvalues of Hamiltonian (2.1). The Algebraic structure of IBM has been described in detail in Refs.[18-21]. Here, we briefly outline the basic ansatz and summarize the results have obtained in this paper for our considered representation. The classification of states in the IBM $SO(6)$ limit is [19-20]

$$U(6) \supset SO(6) \supset SO(5) \supset SO(3) \supset SO(2) \quad , \tag{2.2}$$
$$\downarrow \quad \downarrow \quad \downarrow \quad \downarrow \quad \downarrow$$
$$[N] \quad \langle \Sigma \rangle \quad (\tau) \quad L \quad M$$

The multiplicity label $\nu_\Delta$ in the $SO(5) \supset SO(3)$ reduction will be omitted in the following when it is not needed. The eigenstates $|[N]\langle \sigma \rangle (\tau) \nu_\Delta LM\rangle$ are obtained with a Hamiltonian with the $SO(6)$ dynamical symmetry. The construction of our considered representation requires n-boson creation and annihilation operators with definite tensor character in the basis (2.2) as;

$$B^\dagger_{[N]\langle\sigma\rangle(\tau)lm} \quad , \quad \tilde{B}_{[n^5]\langle\sigma\rangle(\tau)lm} \equiv (-1)^{l-m}(B^\dagger_{[N]\langle\sigma\rangle(\tau)l,-m})^\dagger \quad , \tag{2.3}$$

Of particular interest are tensor operators with $\sigma < n$. They have the property

$$\tilde{B}_{[n^5]\langle\sigma\rangle(\tau)lm}|[N]\langle N\rangle(\tau)\nu_\Delta LM\rangle = 0 \quad , \quad \sigma < n \tag{2.4}$$

For all possible values of $\tau$ and $L$ contained in the $SO(6)$ irrep $\langle N \rangle$. This is so because the action of $\tilde{B}_{[n^5]\langle\sigma\rangle(\tau)lm}$ leads to an $(N-n)$-boson state which contains the $SO(6)$ irrep $\langle \Sigma \rangle = \langle N-n-2i \rangle, i = 0,1,...$, which cannot be coupled with $\langle \sigma \rangle$ to yield $\langle \Sigma \rangle = \langle N \rangle$, since $\sigma < n$. Number conserving normal ordered interactions that are constructed out of such tensors with $\sigma < n$ (and their Hermitian conjugates) thus have $|[N]\langle N\rangle(\tau)\nu_\Delta LM\rangle$ as eigenstates with zero eigenvalues. A systematic enumeration of all interactions with this property is a simple matter of $SO(6)$ coupling. For one body operators,

$$B^\dagger_{[1]\langle 1\rangle(0)00} = s^\dagger \equiv b^\dagger_0 \quad , \quad B^\dagger_{[1]\langle 1\rangle(1)2m} = d^\dagger_m \equiv b^\dagger_{2m} \quad , \tag{2.5}$$

On the other hand, coupled two body operators are of the form

$$B^\dagger_{[2]\langle\sigma\rangle(\tau)lm} \propto \sum_{\tau_k \tau_{k'}} \sum_{kk'} C^{\langle\sigma\rangle(\tau)l}_{\langle l\rangle(\tau_k)k,\langle l\rangle(\tau_{k'})k'} (b^\dagger_k b^\dagger_{k'})^{(l)}_m \quad , \tag{2.6}$$

Where $(b^\dagger_k b^\dagger_{k'})^{(l)}_m$ represent coupling to angular momentum $(l)$ and the $C$ coefficients are known $SO(6) \supset SO(5) \supset SO(3)$ isoscalar factors [21]. These processes lead to the normalized two-boson $SO(6)$ representation displayed in Tables (1-3) for systems with total boson number $N = 2, 3$ and 4, respectively.

Table1. The $SO(6)$ representation of eigenstates for systems with total boson number $N (= n_s + n_d) = 2$.

| $n_d$ | $\sigma$ | $\tau$ | $l$ | Representation |
|---|---|---|---|---|
| 2 | 2 | 2 | 4 | $\sqrt{1/2}(d^\dagger \times d^\dagger)^4_m$ |
| 2 | 2 | 2 | 2 | $\sqrt{1/2}(d^\dagger \times d^\dagger)^2_m$ |
| 2 | 0 | 0 | 0 | $\sqrt{1/2}(d^\dagger \times d^\dagger)^0_0$ |
| 1 | 2 | 1 | 2 | $(s^\dagger \times d^\dagger)^2_m$ |
| 0 | 2 | 0 | 0 | $\sqrt{1/2}(s^\dagger \times s^\dagger)^0_0$ |

Table2. The $SO(6)$ representation of eigenstates for systems with total boson number $N (= n_s + n_d) = 3$.

| $n_d$ | $\sigma$ | $\tau$ | $l$ | Representation |
|---|---|---|---|---|
| 3 | 3 | 3 | 6 | $\sqrt{1/6}[(d^\dagger \times d^\dagger)^4 \times d^\dagger]^6_m$ |
| 3 | 3 | 3 | 4 | $\sqrt{7/22}[(d^\dagger \times d^\dagger)^2 \times d^\dagger]^4_m$ |
| 3 | 3 | 3 | 3 | $\sqrt{7/30}[(d^\dagger \times d^\dagger)^2 \times d^\dagger]^3_m$ |
| 3 | 1 | 1 | 2 | $\sqrt{5/14}[(d^\dagger \times d^\dagger)^0 \times d^\dagger]^2_m$ |
| 3 | 3 | 3 | 0 | $\sqrt{1/6}[(d^\dagger \times d^\dagger)^2 \times d^\dagger]^0_0$ |
| 2 | 3 | 2 | 4 | $\sqrt{1/2}[(d^\dagger \times d^\dagger)^4 \times s^\dagger]^4_m$ |
| 2 | 3 | 2 | 2 | $\sqrt{1/2}[(d^\dagger \times d^\dagger)^2 \times s^\dagger]^2_m$ |
| 2 | 1 | 0 | 0 | $\sqrt{1/2}[(d^\dagger \times d^\dagger)^0 \times s^\dagger]^0_0$ |
| 1 | 3 | 1 | 2 | $\sqrt{1/2}[(d^\dagger \times s^\dagger)^2 \times s^\dagger]^2_m$ |
| 0 | 3 | 0 | 0 | $\sqrt{1/6}[(s^\dagger \times s^\dagger)^0 \times s^\dagger]^0_0$ |

Table3. The $SO(6)$ representation of eigenstates for systems with total boson number $N (= n_s + n_d) = 4$.

| $n_d$ | $\sigma$ | $\tau$ | $l$ | Representation | $n_d$ | $\sigma$ | $\tau$ | $l$ | Representation |
|---|---|---|---|---|---|---|---|---|---|
| 4 | 4 | 4 | 8 | $\sqrt{1/24}[(d^\dagger \times d^\dagger)^4 \times (d^\dagger \times d^\dagger)^4]^8_m$ | 4 | 4 | 4 | 6 | $\sqrt{7/60}[(d^\dagger \times d^\dagger)^4 \times (d^\dagger \times d^\dagger)^2]^6_m$ |
| 4 | 4 | 4 | 5 | $\sqrt{1/12}[(d^\dagger \times d^\dagger)^4 \times (d^\dagger \times d^\dagger)^2]^5_m$ | 4 | 4 | 4 | 4 | $\sqrt{49/664}[(d^\dagger \times d^\dagger)^2 \times (d^\dagger \times d^\dagger)^2]^4_m$ |
| 4 | 2 | 2 | 4 | $\sqrt{5/36}[(d^\dagger \times d^\dagger)^0 \times (d^\dagger \times d^\dagger)^4]^4_m$ | 4 | 2 | 2 | 2 | $\sqrt{5/36}[(d^\dagger \times d^\dagger)^0 \times (d^\dagger \times d^\dagger)^2]^2_m$ |
| 4 | 0 | 0 | 0 | $\sqrt{5/56}[(d^\dagger \times d^\dagger)^0 \times (d^\dagger \times d^\dagger)^0]^0_0$ | 4 | 4 | 4 | 2 | $\sqrt{5/48}[((d^\dagger \times d^\dagger)^2 \times d^\dagger)^0_0 \times d^\dagger]^2_m$ |
| 3 | 4 | 3 | 6 | $\sqrt{1/6}[(d^\dagger \times d^\dagger)^4 \times (d^\dagger \times s^\dagger)^2]^6_m$ | 3 | 4 | 3 | 4 | $\sqrt{7/22}[(d^\dagger \times d^\dagger)^2 \times (d^\dagger \times s^\dagger)^2]^4_m$ |
| 3 | 4 | 3 | 3 | $\sqrt{7/30}[(d^\dagger \times d^\dagger)^2 \times (d^\dagger \times s^\dagger)^2]^3_m$ | 3 | 2 | 1 | 2 | $\sqrt{5/14}[(d^\dagger \times d^\dagger)^0 \times (d^\dagger \times s^\dagger)^2]^2_m$ |
| 3 | 4 | 3 | 0 | $\sqrt{1/6}[((d^\dagger \times d^\dagger)^2 \times d^\dagger)^0_0 \times s^\dagger]^0_0$ | 2 | 4 | 2 | 4 | $\sqrt{1/4}[(d^\dagger \times d^\dagger)^4 \times (s^\dagger \times s^\dagger)^0]^4_m$ |
| 2 | 4 | 2 | 2 | $\sqrt{1/4}[(d^\dagger \times d^\dagger)^2 \times (s^\dagger \times s^\dagger)^0]^2_m$ | 2 | 2 | 0 | 0 | $\sqrt{1/4}[(d^\dagger \times d^\dagger)^0 \times (s^\dagger \times s^\dagger)^2]^0_0$ |
| 1 | 4 | 1 | 2 | $\sqrt{1/6}[(d^\dagger \times s^\dagger)^2 \times (s^\dagger \times s^\dagger)^0]^2_m$ | 0 | 4 | 0 | 0 | $\sqrt{1/24}[(s^\dagger \times s^\dagger)^0 \times (s^\dagger \times s^\dagger)^0]^0_0$ |

With using these eigenstates, one can determine, energy spectra of considered systems as

$$\langle [N]\langle\sigma\rangle(\tau)v_\Delta LM | H | [N]\langle\sigma\rangle(\tau)v_\Delta LM \rangle = \eta n_d + \frac{\eta-1}{N}\varepsilon , \qquad (2.7)$$

Where $\varepsilon$ denote the matrix elements of quadrupole term in Hamiltonian as presented in Tables (4-6) for systems with $N = 2, 3, 4$, respectively.

Table4. The elements of quadrupole operator in Hamiltonian (2.1) for systems with $N = 2$ which determined by states introduced in Table1.

| L | $\varepsilon$ | L | $\varepsilon$ |
|---|---|---|---|
| 0 | $40\chi^2$ | 2 | $\frac{11}{7}\chi^4 - \frac{46}{7}\chi^2 + 16$ |
| 4 | $2 + \frac{18}{7}\chi^2$ | | |

Table5. The elements of quadrupole operator in Hamiltonian (2.1) for systems with $N = 3$ which determined by states introduced in Table2.

| L | $\varepsilon$ | L | $\varepsilon$ |
|---|---|---|---|
| 0 | $315 + 360\chi^2 + \frac{720}{7}\chi^4$ | 3 | $3 + \frac{18}{7}\chi^2$ |
| 2 | $187 + \frac{7113}{7}\chi^2 + \frac{5373}{49}\chi^4 + \frac{473}{343}\chi^6$ | 4 | $33 - 8\chi^2 + \frac{396}{49}\chi^4$ |
| 6 | $3 + \frac{33}{7}\chi^2$ | | |

Table6. The elements of quadrupole operator in Hamiltonian (2.1) for systems with $N = 4$ which determined by states introduced in Table3.

| L | $\varepsilon$ |
|---|---|
| 0 | $31360 + \frac{480454}{7}\chi^2 + \frac{272538}{7}\chi^4 + \frac{69120}{49}\chi^6$ |
| 2 | $402 + \frac{2056}{7}\chi^2 + \frac{3108}{7}\chi^4 + \frac{3971}{49}\chi^6 + \frac{316}{343}\chi^8 + \frac{471}{2401}\chi^{10}$ |
| 3 | $14 + \frac{18}{7}\chi^2$ |
| 4 | $275 + \frac{1902}{7}\chi^2 + \frac{167}{7}\chi^4 + \frac{1059}{49}\chi^6 + \frac{605}{343}\chi^8$ |
| 5 | $4 + \frac{24}{7}\chi^2$ |
| 6 | $56 + \frac{290}{7}\chi^2 + \frac{1221}{49}\chi^4$ |
| 8 | $4 + \frac{52}{7}\chi^2$ |

In our present model, for given $\chi$, there is only one parameter to be determined, namely, the strengths of the $n_d$ term which as have described in the extended Casten triangle, we expect to be $\eta \sim 0$. The energy spectra of considered systems while presented in Figure 2 are determined with using the Eq. (2.7) and matrix elements represented in Tables (4-6). The apparent level-crossing (especially for $\chi = 0$ which correspond to $O(6)$ dynamical symmetry limit) propose a phase transition in this region.

## 2.2. $B(E2)$ Transition

The reduced electric quadrupole transition probabilities $B(E2)$ are considered as observable which as well as quadrupole moment ratios within the low-lying state bands prepare more information about the nuclear structure. The E2 transition operator must be a Hermitian tensor of rank two and also, the number of bosons must be conserved, consequently with these constraints, there are two operators possible in the lowest order. The electric quadrupole transition operator

$$T(E2) = e[(s^\dagger \tilde{d} + d^\dagger s)^{(2)} + \chi (d^\dagger \tilde{d})^{(2)}] \qquad , \qquad (2.8)$$

would employ in the consistent-$Q$ formalism [15-16], namely, with the same $\chi$ value as the Hamiltonian. The reduced electric quadrupole transition rates between $|\alpha_i J_i\rangle \to |\alpha_f J_f\rangle$ states are given by

$$B(E2; \alpha_i J_i \to \alpha_f J_f) = \frac{|\langle \alpha_f J_f \| T(E2) \| \alpha_i J_i \rangle|}{2I_i + 1} \qquad , \qquad (2.9)$$

Where, the matrix elements of the electric quadrupole transition operator defined as

$$\langle \alpha_f J_f \| T^k \| \alpha_i J_i \rangle = (2k+1) \sum_{M_i M_f} (-1)^{J_f - M_f} \begin{pmatrix} J_f & k & J_i \\ -M_f & \kappa & M_i \end{pmatrix} \langle \alpha_f J_f M_f | T_\kappa^k | \alpha_i J_i M_i \rangle \qquad , \qquad (2.10)$$

Now, with using the $SO(6)$ representation of eigenstates and method has been introduced in Refs. [2-4], the quadrupole transition rates ($B(E2)$) in the $SU(3) \leftrightarrow \overline{SU(3)}$ transitional region are determined which displayed in Figure (3). Similar to the energy spectra, the significant variations in $B(E2)$ values, propose a phase transition between these limits. On the other hand, as have predicted in Ref.[16] by Jolie *et al*, the $B(E2; 2_2^+ \to 2_1^+)$ value should peak with a collective value at $O(6)$ and the decrease quickly as $|\chi|$ increases where our results reveal this prediction. The general behavior of energy spectra (the apparent level-crossing) with the $B(E2; 2_2^+ \to 2_1^+)$ values are obvious signs of prolate ($SU(3)$) to oblate ($\overline{SU(3)}$) transition.

## 3. Comparison between theoretical results and empirical data for nuclei with $O(6)$ dynamical symmetry

We applied the above mentioned procedures in the determination of energy spectra and $B(E2)$ transition rates of $^{200-204}_{80}Hg$ isotopes. Theses nuclei have been interpreted as the $O(6)$ like nuclei and have been discussed extensively by the IBM [10,16] and other models [22-28]. These atomic nuclei, as have introduced in Refs.[15-17] and also as have proposed via their $R_{4/2}$ values, seems to be the closet to the phase transition and are here fitted with $\chi = 0$.

## 3.1. Energy spectra

With employing Eq.(2.7) and matrix elements of quadrupole term introduced in Tables (4-6), some low-lying energy levels of considered systems are determined which presented in Table7. A general agreement between the calculation and experimental data is achieved for all considered isotopes.

Table7. Energy levels of considered nuclei which obtained by Eq.(2.7) with $\chi = 0$, empirical data taken from Refs.[29-31].

| L | $^{200}_{80}Hg$ (N=4) | | $^{202}_{80}Hg$ (N=3) | | $^{204}_{80}Hg$ (N=2) | |
|---|---|---|---|---|---|---|
| | Exp. | Calc. | Exp. | Calc. | Exp. | Calc. |
| $2^+_1$ | 367.44 | 391.2 | 439.51 | 401.39 | 436.55 | 440.71 |
| $4^+_1$ | 947.24 | 1002.71 | 1119.84 | 1153.7 | 1128.23 | 1152.91 |
| $0^+_2$ | 1029.34 | 1102.1 | 1411.37 | 1388.61 | 1635.76 | 1655.4 |
| $2^+_2$ | 1254.09 | 1219.06 | 959.94 | 931.44 | 1716.76 | 1691.83 |
| $4^+_2$ | | 1421.43 | 1311.53 | 1371.21 | | 1483.28 |
| $0^+_3$ | 1515.08 | 1498.72 | 1564.78 | 1620.41 | | 1372.52 |
| $2^+_3$ | 1573.66 | 1601.33 | 1182.26 | 1163.72 | 1947.69 | 1920.18 |
| $4^+_3$ | | 1630.12 | 1624.00 | 1598.5 | | 1735.82 |
| $3^+_1$ | 1659.00 | 1741.91 | 1561.98 | 1609.73 | 2094.46 | 2065.44 |
| $2^+_4$ | 1593.42 | 1644.19 | 1389.56 | 1404.38 | 1989.36 | 2023.17 |
| $6^+_1$ | 1706.71 | 1791.62 | 1988.59 | 2026.39 | 2191.01 | 2164.02 |

## 3.2. $B(E2)$ Transition probabilities

The stable even-even nuclei in $Hg$ isotopic chain provide an excellent opportunity for studying the behavior of the total low-lying $E2$ strengths in the $SU(3) \leftrightarrow \overline{SU(3)}$ transitional region. Computation of electromagnetic transition is a sign of good test for the nuclear model wave functions. With using the eigenstates introduced in Tables (1-3) and Eq.(2.8-10), the ratio of different quadrupole transition rates are presented in Table8.

Table8. The ratio of $B(E2)$ transition rates for considered nuclei. Experimental data taken from Refs.[29-31] where determined values obtained via Eq.(2.9) with $\chi = 0$.

| $R_{L+2 \to L}$ | $^{200}_{80}Hg$ | | $^{202}_{80}Hg$ | | $^{204}_{80}Hg$ | |
|---|---|---|---|---|---|---|
| | Exp. | Calc. | Exp. | Calc. | Exp. | Calc. |
| $\dfrac{B(E2; 4^+_1 \to 2^+_1)}{B(E2; 2^+_1 \to 0^+_1)}$ | 1.026 | 1.035 | 2.615 | 2.592 | 1.386 | 1.404 |
| $\dfrac{B(E2; 6^+_1 \to 4^+_1)}{B(E2; 2^+_1 \to 0^+_1)}$ | 1.226 | 1.307 | 1.390 | 1.366 | 1.571 | 1.602 |
| $\dfrac{B(E2; 2^+_2 \to 2^+_1)}{B(E2; 2^+_1 \to 0^+_1)}$ | 0.113 | 0.122 | 0.840 | 0.831 | | 0.958 |

A comparison between the calculated results of the present analysis for different quadrupole transition ratios with experimental data, interprets a satisfactory agreement. In all tables of the present paper, the uncertainties of the experimental data which are smaller than the size of the symbols are not represented.

As it can be seen from these Tables, the calculated energy spectra in this approach are generally in good agreements with the experimental data. The considered results indicate the elegance of the fits presented in this technique and they suggest the success of the guess in parameterization. Also, the calculated $B(E2)$ transition probabilities of $^{200-204}_{80}Hg$ isotopes exhibit nice agreement with experimental ones. These results give information on the structural changes in nuclear deformation and shape-phase transitions. The phase/shape transition was associated with a sudden change in nuclear collective behavior reflected in a sudden increase of $R_{4/2} = E_{4_1^+}/E_{2_1^+}$ from the $O(6)$ dynamical symmetry limit value of 2.5 to the $SU(3)$ dynamical symmetry limit value, 3.3. On the other hand, the significant changes in the $B(E2; 2_2^+ \to 2_1^+)$ values might explain why $^{200-204}_{80}Hg$ isotopes have non- vanishing quadrupole moments although in most other respects they behave as good candidates for $O(6)$ dynamical symmetry limit. As have described in Ref.[15-17], the one-parameter Hamiltonian also explains very well the $R_{4/2}$ ratio on the prolate side of the phase transition, i.e., for negative $\chi$ values. At the phase transition and on the oblate side deviations in the $R_{4/2}$ ratio are observed. In particular, the $^{200-204}_{80}Hg$ isotopes have a slightly smaller $R_{4/2}$ ratio than can be achieved with the simple Hamiltonian. Surprising are the signatures for $^{190-198}_{80}Hg$, while they are not very well described quantitatively by the one-parameter Hamiltonian, they qualitatively reveal unexpected features since they do not resemble a vibrational or shell model structure that might have been expected as $^{208}Pb$ is approached. Such structures would have $R_{4/2}$ around or below 2. Instead a slight increase in $R_{4/2}$ suggests an increase in the deformation which indicates a deviation from the $U(5) \leftrightarrow O(6)$ line towards $SU(3)$. The origin of the increased deformation should be related to the quenching of the pairing correlations at the oblate $Z = 80$ and $N = 120$ subshells [32-33].

## 5. Summary and conclusion

In this paper, the $SU(3) \leftrightarrow \overline{SU(3)}$ transitional region have described by using the $SO(6)$ representation of eigenstates. In the parameter free approach, energy spectra and quadrupole transition rates have determined for systems with total boson number $N = 2$, 3 and 4 while level crossing and significant variation in $B(E2)$ values are apparent. Also, the determined values with $\chi = 0$ for considered quantities, explore satisfactory agreement with empirical data for $^{200-204}_{80}Hg$ isotopes. These results may interpret, $^{200-204}_{80}Hg$ nuclei as situated on or very near the $SU(3) \leftrightarrow \overline{SU(3)}$ leg of the extended Casten triangle. They exhibit a prolate-oblate phase transition in their shape which can be described at the transition by the $O(6)$ limit of the IBM. The results obtained reinforce this new interpretation.

# References

[1]. P. Cejnar, J. Jolie, and R. F. Casten, Rev. Mod. Phys82(2010)2155.
[2]. D.J. Rowe,Nucl.PhysA745 (2004) 47.
[3]. G. Rosensteel a, D.J. Rowe,Nucl.PhysA759 (2005) 92.
[4]. D.J. Rowe,G.Thiamova.Nucl.PhysA760 (2005) 59.
[5]. F. Iachello, Phys. Rev. Lett85(2000)3580.
[6]. F. Iachello, Phys. Rev. Lett87(2001)052502.
[7]. F. Iachello , Phys.Rev.Lett91(2003)132502.
[8]. Dennis Bonatsos,D. Lenis,D. Petrellis,P.A. Terziev, Phys.Lett.B588(2004)172.
[9]. F. Iachello and A. Arima, The Interacting Boson Model (Cambridge University Press, Cambridge, England, 1987).
[10]. F. Iachello, A. Arima, Ann. Phys. (N.Y.) 115(1978)325 and 99(1976)253 and 111 (1978)201 and 123(1979)468.
[11]. O. Scholten , A. E. L. Dieperink and F. Iachello, Phys.Rev. Lett44(1980)1747.
[12]. R.F.Casten, D.D.Warner, Rev. Mod. Phys60(1988)389.
[13]. E. Lopez-Moreno and O. Castanos, Phys. Rev. C 54(1996)2374.
[14]. E. Lopez-Moreno and O. Castanos, Rev. Mex. Fis.44(1998)48.
[15]. J. Jolie, R. F. Casten, P. von Brentano, and V. Werner,Phys.Rev.Lett87(2001)162501.
[16]. J. Jolie, A. Linnemann,Phys.RevC 68(2003) 031301.
[17]. G.Thiamova, P. Cejnar.Nucl.PhysA765(2006)97.
[18]. W.Pfeifer, An Introduction to the Interacting Boson Model of the Atomic Nucleus, Part I(nucl-th/0209039).
[19]. A.Leviata, Rev.Mex.Fis42(1996)152.
[20]. J.E.Garcia.Ramos, A.Leviata,P.Van.Isacker, Phys.Rev.Lett102(2009)112502.
[21]. A.Frank,P.Van.Isacker, Algebraic methods in Molecular and Nuclear physics(Wiely New York 1994).
[22]. Yan-An Luo, Jin-Quan Chen, J.P. Draayer, Nucl. Phys.A 669(2000)101.
[23]. T. Mizusaki, T. Otsuka, Prog. Theor. Phys. Suppl. 125(1996)97.
[24]. X.W. Pan, J.L. Ping, D.H. Feng, J.Q. Chen, C.L. Wu, M. Guidry, Phys. Rev. C 53(1996)715.
[25]. X.W. Pan, D.H. Feng, Phys. Rev. C 50(1994)818.
[26]. A. Leviatan , Phys.Rev.Lett98(2007)242502.
[27]. V.Werner, P.von Brentano, R.F. Casten, J.Jolie, Phys.Lett.B527(2002)55.
[28]. ZHANG JinFu,BAI HongBo, Chinese Science Bulletin52(2007) 165.
[29]. National Nuclear Data Center,(Brookhaven National laboratory), chart of nuclides, (http://www.nndc.bnl.gov/chart/reColor.jsp?newColor=dm)
[30]. Live chart, Table of Nuclides, (http://www-nds.iaea.org/relnsd/vcharthtml/VChartHTML.html).
[31]. Richard B. Firestone,Virginia S. Shirley, S. Y. Frank , Coral M. Baglin and Jean Zipkin, table of isotopes(1996).
[32]. M. Vergnes, G. Berrier-Ronsin, G. Rotbard, J. Skalski, and W.Nazarewicz, Nucl. Phys. A514(1990)381.
[33]. D.D. Warner, Nature(London)420(2002)614.

# Figure Caption

**Figure1.** The extend Casten triangle [12.14], represents different dynamical symmetries of IBM as open circles.

**Figure2.** Energy spectra for systems with $N=4$, determined via Eq.(2.7) and matrix elements introduced in Table 6. The apparent level crossing in this region, suggest a phase transition.

**Figure3.** The variation of $B(E2)$ values in the transitional region for systems with $N=4$, determined via Eq.(2.8-10). Similar to energy spectra, significant variations of $B(E2)$ values, propose a phase transition between these limits.

Figure1.

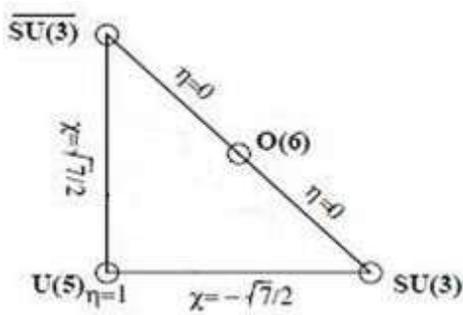

Figure2.

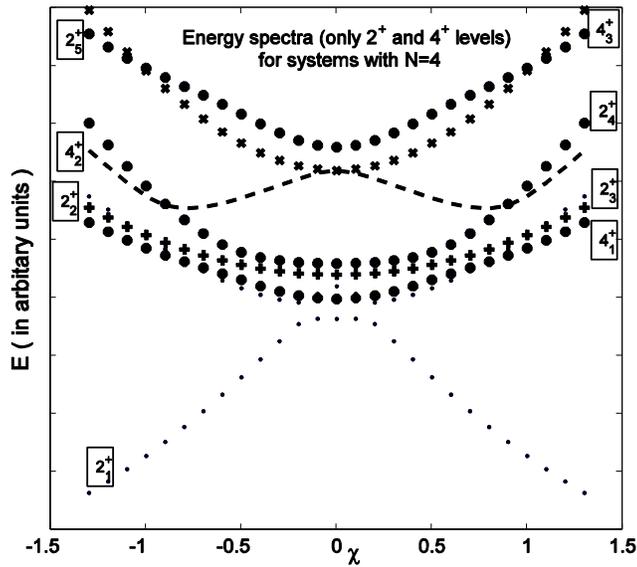

Figure3.

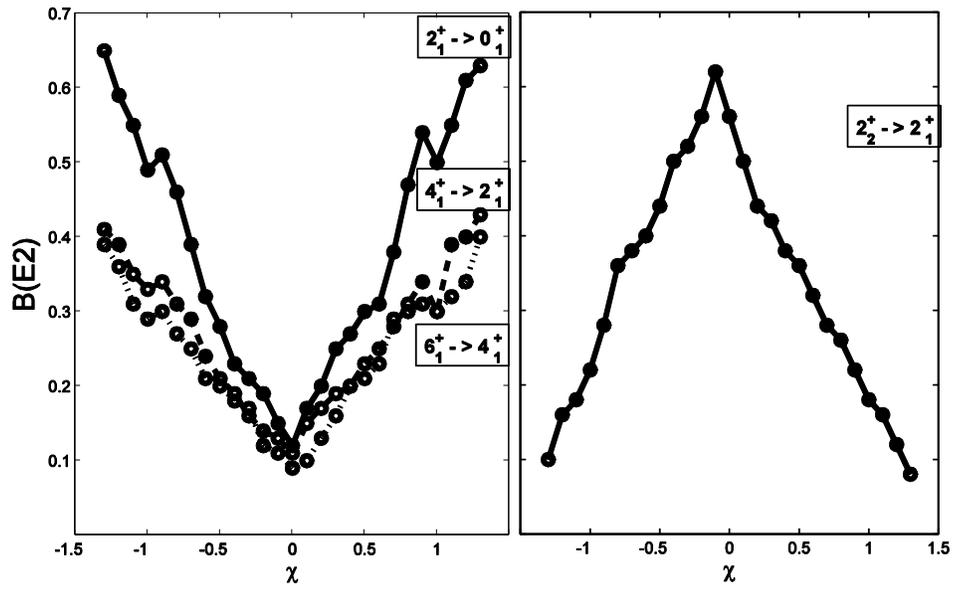